\documentclass{aastex}
\usepackage{emulateapj5}

\renewcommand{\tabcolsep}{3pt}

\newcommand{\NH}{{$N_{\rm H}$}}

\newcommand{\eps}{ergs s$^{-1}$}
\newcommand{\pcm}{cm$^{-2}$}
\newcommand{\ps}{s$^{-1}$}
\newcommand{\chandra}{{\it Chandra}}
\newcommand{\asca}{{\it ASCA}}
\newcommand{\rosat}{{\it ROSAT}}
\newcommand{\Einstein}{{\it Einstein}}
\newcommand{\Ginga}{{\it Ginga}}

\slugcomment{Accepted for publication in {\it The Astrophysical Journal}.}

\lefthead{Terashima et al.}
\righthead{Chandra Observation of M51}

\begin{document}

\title{A Chandra Observation of M51: Active Nucleus and
Nuclear Outflows}

\author{Yuichi Terashima\altaffilmark{1, 2} and
Andrew S. Wilson\altaffilmark{2, 3}
}

\altaffiltext{1}{NASA Goddard Space Flight Center, Code 662,Greenbelt, MD 20771}

\altaffiltext{2}{Astronomy Department, University of Maryland, College Park, 
MD 20742}

\altaffiltext{3}{Adjunct Astronomer, Space Telescope Science Institute, 3700
San Martin Drive, Baltimore, MD 21218}

\begin{abstract}

We present a {\chandra} ACIS-S observation of the nuclear region of
the nearby spiral galaxy M51 (NGC 5194), which has a low-luminosity
Seyfert 2 nucleus.  The X-ray image shows the nucleus, southern
extranuclear cloud, and northern loop, the morphology of the extended
emission being very similar to those seen in radio continuum and
optical emission line images.  The X-ray spectrum of the nucleus is
well represented by a model consisting of soft thermal plasma with $kT
\approx 0.5$ keV, a very hard continuum, and an Fe K$\alpha$ emission
line at 6.45 keV with an equivalent width of $>$ 2 keV. The very
strong Fe line and the flat continuum indicate that the nucleus is
obscured by a column density in excess of $10^{24}$ {\pcm} and the
spectrum is dominated by reflected emission from cold matter
near the nucleus.  The X-ray spectra of the extranuclear clouds
are well fitted by a thermal plasma model with $kT\approx 0.5$
keV. This spectral shape and morphology strongly suggest that the
clouds are shock heated by the bi-polar outflow from the nucleus. The
shock velocities of the extranuclear cloud and northern loop inferred
from the temperatures of the X-ray gas are 690 km {\ps} and 660 km
{\ps}, respectively.
By assuming a steady-state situation in which the emission of the
extranuclear clouds is powered by the jets, the mechanical energy in
the jets is found to be comparable to the bolometric luminosity of the
nucleus.

\end{abstract}

\keywords{galaxies: active --- galaxies: individual (M51) --- galaxies: 
nuclei --- galaxies: Seyfert --- X-rays: galaxies --- shock waves}

\section{Introduction}

M51 (NGC 5194) is a nearby (distance 8.4 Mpc; Feldmeier et al. 1997)
spiral galaxy hosting a low-luminosity active galactic nucleus (AGN).
The nucleus shows optical emission lines and is classified as a Seyfert 2
(e.g., Ford et al. 1985; Ho, Filippenko, \& Sargent 1997).  Radio
continuum and optical emission-line observations reveal a bi-polar
structure comprising a southern extranuclear cloud (hereafter XNC) and
a northern loop; this morphology and the high gas velocities observed
are indicative of nuclear outflow
(Ford et al. 1985, Cecil 1988, Crane \& van der Hulst
1992). A radio jet connecting the nucleus and the XNC is also seen in
a 6 cm image (Crane \& van der Hulst 1992). The optical emission-line
clouds are inferred to be heated by the jet emanating from the nucleus.

X-ray observations are a powerful tool to investigate the physical
nature of these phenomena. However, the spatial resolution of previous
observations is not good enough to resolve the structures in the
nuclear region.  In the soft X-ray regime, extended nuclear emission
was detected with the {\Einstein} HRI (Palumbo et al. 1985), the
{\rosat} PSPC (Marston et al. 1995), and the {\rosat} HRI (Ehle,
Pietsch, \& Beck 1995; Roberts \& Warwick 2000), but only a hint of
enhanced emission in the direction towards the XNC is seen in the
{\rosat} HRI data.  The spectrum of the nuclear region obtained with
the {\rosat} PSPC is very soft.  Hard X-ray emission,with an X-ray
luminosity of 9$\times10^{40}$ {\eps} (2--20 keV) was detected by
{\Ginga} and interpreted as a combination of low-mass X-ray binaries
in the host galaxy and a highly absorbed AGN (Makishima et al. 1990).
Subsequently, {\asca} detected a strong Fe K$\alpha$ emission line
with an equivalent width (EW) of $\approx$ 1 keV, which probably comes
from the heavily obscured AGN, as well as emission from thermal plasma
with a temperature of $\approx$0.5 keV. The 2--10 keV luminosity of
the entire galaxy measured by {\asca} was $8.4\times10^{39}$ {\eps}
which is about six times smaller than that obtained with {\Ginga}
in the same band (Terashima et al. 1998). The reason for this
large difference in luminosities has recently been clarified by {\it
BeppoSAX} observations. Fukazawa et al. (2001) find that the nucleus
is photoelectrically absorbed below 10 keV, but seen directly through
the absorption at 20 keV, implying an absorbing column of {\NH} =
$5.6\times10^{24}$ {\pcm}. The 2--10 keV luminosity measured by {\it
BeppoSAX} was similar to that of {\asca}. Fukazawa et al. (2001)
attributed the higher {\Ginga} luminosity to variability of the
absorbing column, as long as the {\Ginga} measurement is not contaminated
by nearby bright sources.  None of these X-ray studies has sufficient
spatial resolution to resolve the nucleus and outflows seen in radio
continuum and optical [\ion{N}{2}]$\lambda\lambda$6548,6583 images.


High spatial resolution X-ray observations with the {\chandra} X-ray
observatory can easily resolve the multiple components in the nuclear
region and enable us to study the physical nature of the nucleus and
outflows, as well as the various X-ray sources in the host galaxy.  In
this {\it Letter}, we present {\chandra} results on the nuclear region
of NGC 5194. Results on the diffuse X-ray emission and discrete X-ray
sources seen in the host galaxy will be presented in a future paper. At the
distance of 8.4 Mpc, 1 arcsec corresponds to 40.7 pc.

\newpage

\section{Observation and Data Analysis}

M51 was observed on 2000 June 20 with the {\chandra} Advanced CCD
Imaging Spectrometer (ACIS; Garmire et al. 2001, in preparation). M51
was located on the back illuminated CCD chip S3. The data were reduced
using the CIAO software version 1.1.5. Gain correction was made using
the latest calibration files. We used only {\it ASCA} event grades 0,
2, 3, 4, and 6 (see {\chandra} Proposer's Observatory Guide,
2000). The background was stable during the observation and a net
exposure time of 14865 sec was obtained. The count rates of all the
detected sources in the field were relatively low and effects of
pileup were negligible. Spectral fits were performed using XSPEC
version 11.0 and all the quoted errors are 90\% confidence range for
one parameter of interest.

\section{Results and Discussion}

\subsection{X-ray Images}

Our {\chandra} image of NGC 5194 shows X-rays from a bright nuclear
region, diffuse emission distributed over the host galaxy, and many
discrete sources. The companion galaxy NGC 5195 is also detected in the
same field. In this {\it Letter}, we concentrate on the nuclear region
of NGC 5194.



An expanded image of the central region of NGC 5194 in the 0.5--8 keV
band is shown in Figs 1 and 2. In Fig. 1, red, green, and blue colors
correspond to the energy bands 0.5--1.3, 1.3--3, and 3--8 keV,
respectively. The nucleus, extended features, and a few discrete
sources are seen. The extended structures north and south of the
nucleus are remarkably similar to the features seen in radio (6 cm ,
20 cm) and [\ion{N}{2}] images, and we refer to them as the northern
loop and the XNC, respectively. Fig. 2 shows a 20 cm image (contours,
from Ford et al. 1985) overlaid on a {\chandra} image in the 0.5--8
keV band.  Both images have $\approx 1^{\prime\prime}$ resolution and
a close correspondence is seen between the two maps. The nuclei
were aligned by shifting the X-ray image by 0.4$^{\prime\prime}$.

The color (=hardness) map (Fig. 1) shows that the nucleus has a hard
spectrum while the XNC and the northern loop are soft. A spectral
analysis (next subsection) shows that the nucleus emits both hard and
soft spectral components, the latter being very similar to the spectra
of the XNC and northern loop. The north-east part of the XNC is
somewhat harder than the rest of the XNC and the northern loop. It is
intriguing to note that this region coincides with a bright knot in
the [\ion{N}{2}] image from which the broadest optical emission lines
(extending over $\sim1500$ km {\ps}) are found (the point P in Cecil
1988). It is also worth noting that the elongation of the nucleus seen
in [\ion{N}{2}] is towards the point P, and that the radio jet
connects the nucleus and point P (Crane \& van der Hulst 1992). Thus
this medium-hardness region might be the site where the jet emanating
from the nucleus is currently interacting with the interstellar medium
and shock heating it.

\subsection{X-ray spectra}


\subsubsection{Nucleus}

The spectrum of the nucleus was extracted from a circular region with
a radius of 1.5 arcsec. The spectrum was binned so that each bin
contains at least three counts in order to enhance the visibility of
the narrow Fe-K$\alpha$ emission line. In this case, a chi-squared fit
is not adequate and a maximum-likelihood method using the C-statistic
(Cash 1979) was employed.  In the fit with the C-statistic, background
cannot be subtracted, so we first constructed a model of the
background spectrum.  This background model was obtained from an area
away from the nucleus, which includes diffuse emission in the host
galaxy. The spectrum from this area was fitted by a two temperature
($kT$=0.23 keV, 0.60 keV; abundance 0.16 solar) plasma model plus a
flat power law. This fixed background model was then added to the
model for the nucleus after normalizing by the geometrical area.

The spectrum of the nucleus (Fig. 3a) consists of at least three
components, namely a soft component showing emission lines from
ionized species, a hard component which dominates the 3--8 keV band,
and a strong emission line at 6.4 keV. We modeled the soft component
with the optically-thin thermal plasma model MEKAL. Two models were
examined for the hard component: an absorbed power law or pure
reflection from almost face-on cold matter (Magdziarz \& Zdziarski
1995). The latter corresponds to a situation in which the X-ray emission
from the nucleus is almost completely obscured by a column density
in excess of $10^{24}$ {\pcm} and only emission scattered by matter
near the nucleus is observed.  We added a Gaussian component to
represent the emission line at 6.4 keV.  The absorbed power law and
pure reflection model provide a similar quality of fit (C-statistic =
39.1 and 39.6, respectively).  The spectral parameters obtained are
summarized in Table 1.

In the case of the best-fit absorbed power law model, we fixed the
photon index of the power law component at 2.0 (typical of unobscured
AGN) and obtained an absorption column density of
{\NH}=$3.1^{+1.8}_{-1.5}\times10^{23}$ {\pcm}.  The emission line
centroid energy and equivalent width (EW) are
$6.458^{+0.033}_{-0.039}$ keV (source rest frame) and
$3.5^{+2.7}_{-1.6}$ keV, respectively. Although this model can
describe the data, the very large EW of the Fe line is not compatible
with the relatively small absorption column ($3\times10^{23}$ {\pcm}),
from which an Fe line with much lower EW ($\sim 200$ eV) is expected
(e.g, Awaki et al. 1991; Leahy \& Creighton 1993; Ghisellini, Haardt,
\& Matt 1994 ).

The center energy, line width, and EW of the strong emission line in
the pure reflection model are $6.458^{+0.035}_{-0.048}$ keV at the
source rest frame, $\sigma$ = 0.020 ($<$0.099) keV, and
$4.8^{+4.3}_{-2.5}$ keV, respectively. This line centroid energy is
consistent with fluorescent iron emission from neutral or
low-ionization state ($<$ FeXIX) iron when possible small calibration
uncertainties are taken into account. The line intensity of
$(5.4^{+3.5}_{-2.5})\times10^{-6}$ photons {\pcm} {\ps} is consistent
with that obtained with an {\asca} observation (Terashima et
al. 1998), although the error ranges are large.  The huge EW ($>$2.4
keV) is difficult to explain even in the pure reflection picture and
probably an overabundance of iron is required (Matt, Fabian, \&
Reynolds 1995). Such a large equivalent width is also seen in other
Compton thick Seyfert 2s such as the Circinus galaxy (Matt et
al. 1996). The X-ray luminosity of the reflection component in M51 is
$9.8\times10^{38}$ {\eps} in the 2--10 keV band. Although this
luminosity is nine times smaller than the total 2--10 keV luminosity
of M51 observed by {\asca}, X-ray binaries contribute significantly to
the {\asca} luminosity (Terashima et al. 1998). Therefore, the smaller
luminosity obtained with {\chandra} does not imply time variability of
the AGN. Recently {\it BeppoSAX} observations have shown that
the nucleus is heavily absorbed by a column density  of {\NH} =
$5.6\times10^{24}$ {\pcm} and the intrinsic luminosity corrected for
the absorption is $1.1\times10^{41}$ {\eps} in the
2--10 keV band (Fukazawa et al. 2001). Thus, the luminosity obtained
with {\chandra} is about 1\% of the intrinsic X-ray
luminosity, implying that $\approx$1\% of the nuclear luminosity is
scattered in our direction.

The soft component (Fig. 3a) is well represented by a MEKAL plasma
model (Table 2). The obtained absorption column density is only
slightly larger than the Galactic value ($1.3\times10^{20}$ {\pcm},
Stark et al. 1992) given the large errors. The emission measure is
$1.2\times10^{62}$ cm$^{-3}$ and the luminosity corrected for the
absorption is $4.5\times10^{38}$ {\eps} in the 0.5--4 keV band. The
temperature and abundance are quite similar to the extended emission
(the XNC and north loop).  The gas may be shock heated as discussed
below.

\subsubsection{Southern Extranuclear Cloud and Northern Loop}

We extracted X-ray spectra of the southern extranuclear cloud (XNC)
and the northern loop. A background spectrum was obtained from a
region free of the nuclear emission.  Bright point sources in this
region were excluded. This background spectrum was then subtracted
from the spectra of the XNC and northern loop. The spectra were binned
so that at least 15 counts were contained in each bin, so chi-squared
statistics can be used.

The two spectra (Fig. 3b and c) show emission lines which are
indicative of optically thin thermal plasma emission with a
temperature below 1 keV. Therefore, we fitted the spectra with the
MEKAL model modified by photoelectric absorption along the line of
sight. This model provides successful descriptions of both the XNC and
the northern loop.  The results of the spectral modeling are shown in
Table 2.

The spectral parameters obtained for the XNC are $kT =
0.58^{+0.04}_{-0.08}$ keV, abundance $0.14^{+0.07}_{-0.05}$ solar, and
{\NH} = $9.5^{+5.5}_{-3.9}\times10^{20}$ {\pcm}. We note that the
abundance values in all the fits in this paper are unusually low for
the central regions of spiral galaxies and are similar to other X-ray
measurements of abundance of spiral galaxies and starburst galaxies.
The reason for these unphysically low abundances is unknown, but may
involve systematic errors in modeling thermal plasmas, nonthermal
continuum contributions, or depletion of gas phase elements onto
grains (Netzer \& Turner 1997, Arimoto et al. 1997, Dahlem et
al. 2000, Matsushita et al. 2000) and does not necessarily mean real
low abundance. The luminosity (corrected for absorption) and emission
measure are $1.5\times10^{39}$ {\eps} in the 0.5--4 keV band and
$2.7\times10^{62}$ cm$^{-3}$, respectively. If we assume a simple
spherical shape with a radius of 80 pc and a filling factor $\eta
(<1)$ of the X-ray emitting gas, we can derive some physical
quantities.  The density $n$, gas mass $M$, and the thermal energy $U$
are determined as $n=2.1\eta^{-1/2}$ cm$^{-3}$,
$M=1.5\times10^5\eta^{1/2}M_{\odot}$, and
$U=1.2\times10^{53}\eta^{1/2}$ ergs, respectively.


This thermal plasma is expected from shock heating due to
outflows. The expected postshock temperature is $T_{\rm s} = 1.4
\times 10^5 ~V_{\rm s7}^2$ K, where $V_{\rm s7}$ is the shock velocity
in units of 100 km s$^{-1}$, if the preshock medium is fully ionized
and the helium abundance is 10\% of hydrogen by number (Hollenbach and
McKee 1979).  Using this relation, the shock velocity is determined as
690 km s$^{-1}$ from the observed temperature of the XNC. This shock
velocity is roughly consistent with that determined from [\ion{N}{2}]
observations (500 km s$^{-1}$; Cecil 1988). The XNC shows LINER-like
optical emission lines (Ford et al. 1985). The emission line ratios
[\ion{O}{1}]/H$\alpha$, [\ion{O}{3}]/H$\beta$, and
[\ion{S}{2}]/H$\alpha$ are in the range expected from fast shocks
without precursors for a shock velocity of 300--500 km {\ps}, as
calculated by Dopita \& Sutherland (1995, 1996). The observed
[\ion{N}{2}]/H$\alpha$ ratio is larger than predicted by the shock
model and could be due to an overabundance of nitrogen, as often
suggested previously. Thus the optical emission line ratios are
consistent with fast shocks and reinforce a shock interpretation of
the origin of the extranuclear clouds.

The spectral parameters for the northern loop are $kT =
0.53^{+0.05}_{-0.04}$ keV, abundance $0.14^{+0.07}_{-0.04}$ solar, and {\NH}
= $6.2^{+2.7}_{-3.3}\times10^{20}$ {\pcm}. The luminosity (corrected
for absorption) and emission measure are $2.0\times10^{39}$ {\eps} in
the 0.5--4 keV band and $3.8\times10^{62}$ cm$^{-3}$, respectively. As
in the case of the XNC, we determine physical parameters of the X-ray
gas by assuming a simple geometry. Here we assume a spherical shell
with an outer radius of 200 pc and a thickness of 40 pc, and a volume
filling factor of $\eta$. The parameters obtained for the hot gas are
$n=0.88\eta^{-1/2}$ cm$^{-3}$, $M=5.0\times10^5\eta^{1/2}M_{\odot}$,
and $U=3.7\times10^{53}\eta^{1/2}$ ergs.

The X-ray spectral shape (i.e. the model's $kT$ and abundance), X-ray
morphology, and optical emission line properties of the northern loop
are quite similar to the XNC and suggest a similar origin to the XNC,
i.e., shock heating by nuclear outflows. The shock velocity required
to heat the gas to $kT$ =0.53 keV is 660 km s$^{-1}$, almost the same
shock velocity as obtained for the XNC. The fact that the northern
loop is much larger than the XNC suggests that it is an older
structure.

The total radiative luminosity of the shock heated gas ($L_{\rm T}$)
can be estimated from the H$\beta$ luminosities (Ford et al. 1985) by
assuming the shock model (Dopita \& Sutherland 1995, 1996). A value
$L_{\rm T} \sim 2\times10^{42}$ {\eps} is obtained. This luminosity is
comparable to the photon luminosity of the nucleus if an intrinsic
X-ray luminosity of $10^{41}$ {\eps} (Fukazawa et al. 2001) and a
bolometric correction of a factor of ten (Ho 1999) are assumed.  If
the shock-heated gas is, indeed, powered by the jets, which are
assumed to be steady, the mechanical and radiative outputs of the
nucleus are then similar to within the considerable uncertainties.

\section{Concluding Remarks}

Our {\chandra} observations have shown that the X-ray morphology of
the extranuclear regions of NGC 5194 is very similar to that seen in
radio continuum and optical emission-line imaging. The X-ray spectra
of the extranuclear cloud and the northern loop are well described by
thermal plasma models with $kT\approx$ 0.5--0.6 keV. This temperature is
close to that expected from shock heating given the range of
velocities seen ($\sim$500 km {\ps}) in optical emission-line
spectra. The optical emission-line ratios also agree with models
involving shocks with velocities in the range 300--500 km {\ps}.
Further, the location with the hardest X-ray spectrum in the
extranuclear cloud coincides with the region exhibiting the broadest
optical lines ($\sim1500$ km {\ps}). This location is also apparently
the terminus of the radio jet. All of these findings indicate that the
extranuclear gas is shock heated by nuclear jets and/or other
collimated outflows. 

  The nucleus itself exhibits a soft X-ray spectrum which is very
similar to that of the extranuclear gas. We presume that this gas is
also shock heated by mass outflows. In addition, the nucleus shows a
very hard continuum and an iron line with equivalent width in excess
of 2 keV. This spectrum arises through reflection of the radiation
from a compact nuclear source with a power law spectrum by cold matter
in its vicinity. The nuclear source itself is obscured by a large
column density of gas ({\NH}=$5.6\times10^{24}$ {\pcm}) and is only
seen directly above 10 keV (Fukazawa et al. 2001).

\acknowledgments

We are grateful to Gerald Cecil for providing the radio and
[\ion{N}{2}] images in computer readable format.  YT was supported in
part by a Japan Society for the Promotion of Science Postdoctoral
Fellowship for Research Abroad. This research was supported by NASA
grant NAG 81027.

\newpage


\begin{table*}
\begin{center}
\caption{Spectral modeling of the hard component of the nucleus.}
\begin{tabular}{cccccccc}
\tableline \tableline
Model	& $N_{\rm H}$	& photon index	& $E_{\rm line}^{a}$ & Width & EW	& Luminosity$^b$ & C-statistic\\
	& [$10^{22}$ cm$^{-2}$] & & [keV]	& [keV]	& [keV]	& [$10^{39}$ ergs s$^{-1}$]	& \\
\hline
1	& $31^{+18}_{-15}$	& 2.0$^c$ & $6.458^{+0.033}_{-0.039}$	& $0(<0.085)$	& $3.5^{+2.7}_{-1.6}$	& 1.1	& 39.1\\
2	&  $0.25(<34)$	& 2.0$^c$ & $6.458^{+0.035}_{-0.048}$	& $0.020(<0.099)$ & $4.8^{+4.3}_{-2.5}$ & 0.98	& 39.5\\
\tableline
\end{tabular}
\end{center}
\tablecomments{Model 1: MEKAL + absorbed power law + Gaussian line, 
Model 2: MEKAL + pure reflection + Gaussian line.
{\it a}: Source rest frame. {\it b}: Observed luminosity (not corrected for absorption) in the 2--10 keV band. {\it c}: Fixed parameter.
}
\end{table*}

\begin{table*}
\begin{center}
\caption{Spectral modeling of the soft thermal component.}
\begin{tabular}{cccccc}
\hline \hline
	& $N_{\rm H}$	& $kT$	& abundance	& Luminosity$^a$	& $\chi^2$/dof\\
	& [$10^{22}$ cm$^{-2}$] & [keV]	& [solar] & [$10^{39}$ ergs s$^{-1}]$	& \\
\hline
Nucleus$^b$	& $0.14^{+0.17}_{-0.11}$ & $0.49^{+0.13}_{-0.08}$	& $0.084\pm0.05$	& 0.45	& ...\\
XNC	& $0.095^{+0.055}_{-0.039}$ & $0.58^{+0.04}_{-0.08}$ & $0.14^{+0.07}_{-0.05}$	& 1.5	& 53.9/47\\
Northern Loop	& $0.062^{+0.027}_{-0.033}$ &$0.53^{+0.05}_{-0.04}$ & $0.14^{+0.07}_{-0.04}$ & 2.0	& 44.9/50\\
\hline
\end{tabular}
\end{center}
\tablecomments{ {\it a}: Luminosity corrected for absorption in the 0.5--4 keV band. {\it b}: The hard component is modeled by a pure reflection continuum + Gaussian line.
}
\end{table*}

\newpage

\clearpage


\begin{figure}[h]
\includegraphics[scale=0.87]{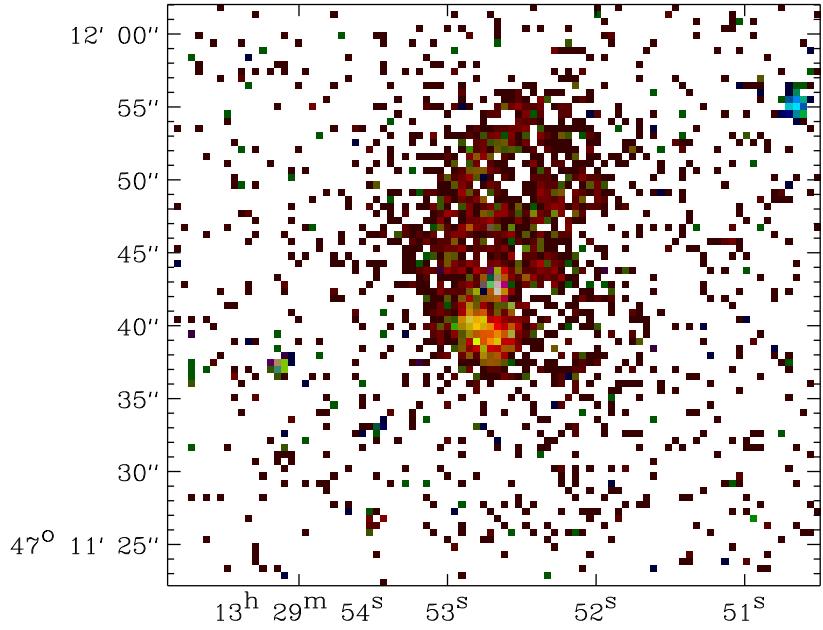}
\caption{{\chandra} ACIS-S image of the central region of NGC
5194.  Red, green, and blue colors correspond to the 0.5--1.3, 1.3--3,
and 3--8 keV band, respectively. The color intensity is scaled
logarithmically to the number of photons in each pixel. The position 
of the optical nucleus is (13$^{\rm h}$29$^{\rm m}$52$^{\rm s}$.7, 47$^{\circ}$11$^\prime$43$^{\prime\prime}$) (J2000).
The size of
the image is 40 arcsec $\times$ 45 arcsec. North is up and east is to
the left.
\label{fig-1}}
\end{figure}


\begin{figure}[h]
\hspace{2.1cm}
\includegraphics[scale=0.49,angle=-90]{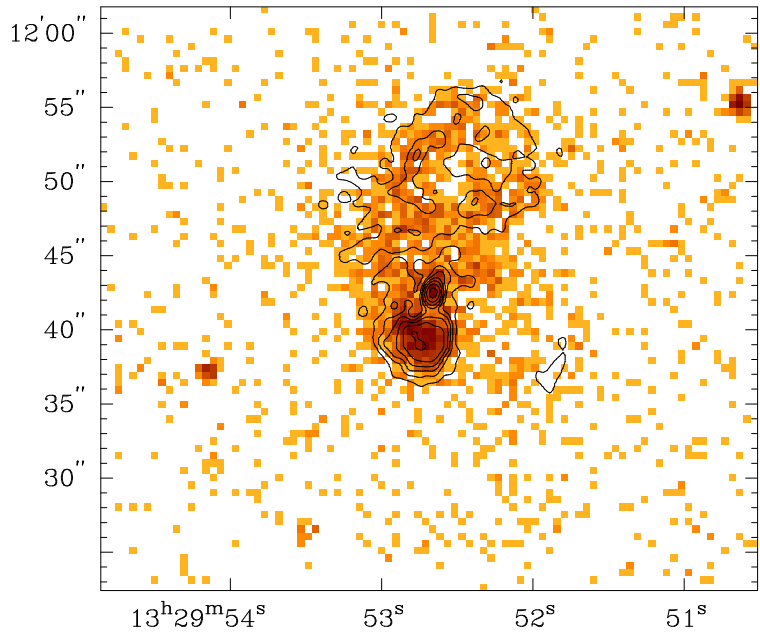}
\figcaption[]{ A VLA 20 cm map (contours, from Ford et al. 1985)
overlaid on a {\chandra} image in the 0.5--8 keV band.  
Contours are plotted at 7.5, 15, 21, 30, 43, 60, 85\% of peak
and the peak corresponds to 2.12 mJy beam$^{-1}$.
\label{fig-2}}
\end{figure}

\begin{figure}[ht]
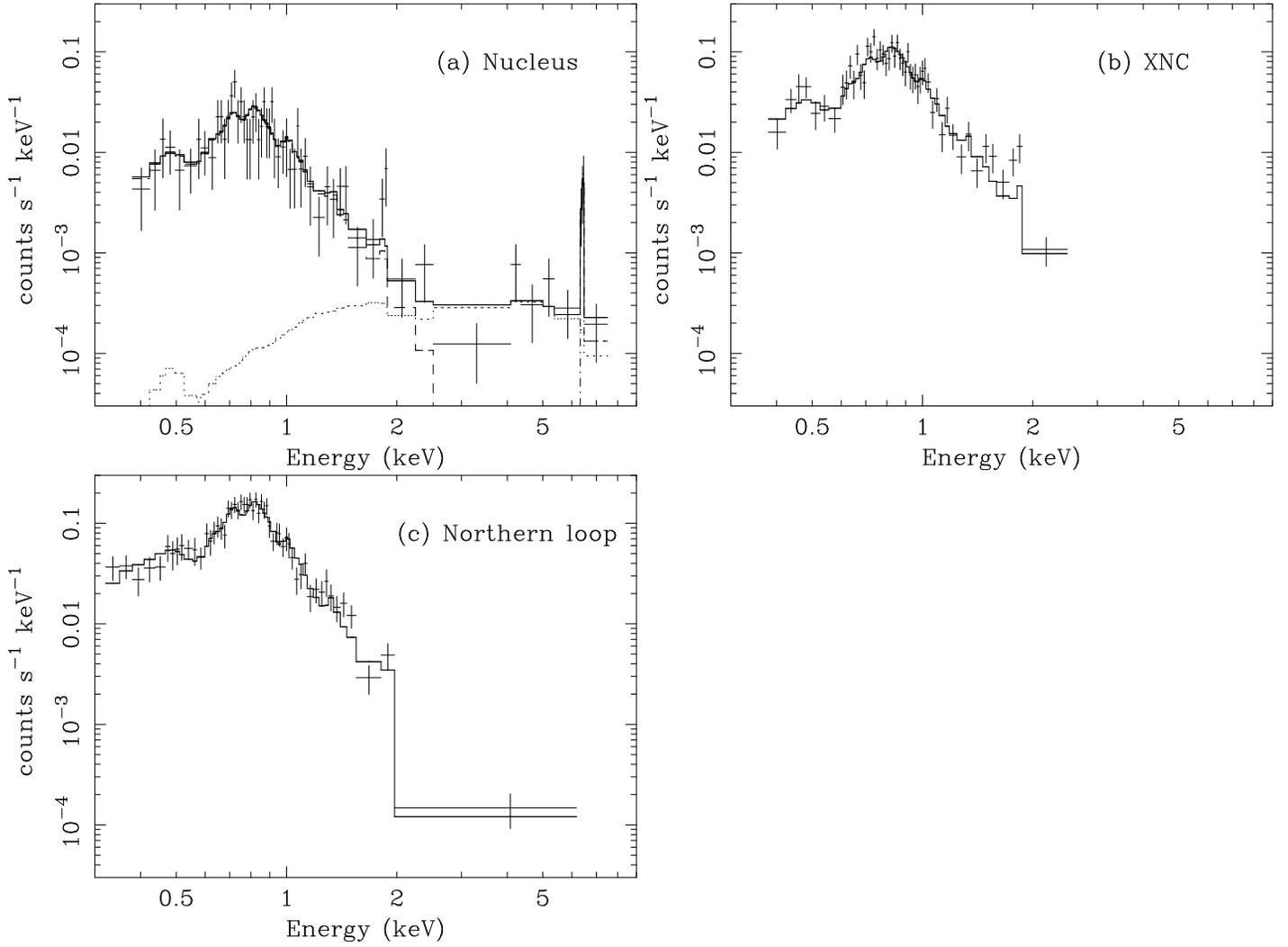

\includegraphics[scale=0.5,angle=-90]{fig3a.ps}
\includegraphics[scale=0.5,angle=-90]{fig3b.ps}
\includegraphics[scale=0.5,angle=-90]{fig3c.ps}
\figcaption[]{X-ray spectra of the nucleus ({\it a}), southern
extranuclear cloud (XNC) ({\it b}), and northern loop ({\it
c}). Crosses are data and histograms are models. The hard component of
the nucleus is assumed to be reflected emission from cold matter
(dotted line), while the soft component of all the three spectra is
assumed to be thermal emission from a gas in collisional-ionization
equilibrium, modeled by the MEKAL code.
\label{fig-3}}
\end{figure}

\end{document}